\documentclass[twocolumn,pra,superscriptaddress,noeprint]{revtex4-1}
\usepackage{graphicx}
\usepackage{amssymb,amsmath}
\usepackage{bm}
\usepackage{dcolumn}
\usepackage{subfigure}
\usepackage{float}
\usepackage{url}
\usepackage{xcolor}
\usepackage{ulem}
\usepackage{sidecap}
\usepackage{bbold}
\allowdisplaybreaks
\usepackage{hyperref}
\hypersetup{backref,pdfpagemode=FullScreen,colorlinks=true,breaklinks,urlcolor=blue,linkcolor=blue,citecolor=blue}

\usepackage{mathrsfs}

\begin{document}

\title{Resonant Spin Exchange between Heteronuclear Atoms Assisted by Periodic Driving}

\author{Jun-Jie Chen}
\affiliation{State Key Laboratory of Low Dimensional Quantum Physics, Department of Physics, Tsinghua University, Beijing 100084, China}
\affiliation{Department of Physics and Shenzhen Institute for Quantum Science and Engineering, Southern University of Science and Technology, Shenzhen 518055, China}

\author{Zhi-Fang Xu}
\email{xuzf@sustc.edu.cn}
\affiliation{Department of Physics and Shenzhen Institute for Quantum Science and Engineering, Southern University of Science and Technology, Shenzhen 518055, China}

\author{Li You}
\email{lyou@mail.tsinghua.edu.cn}
\affiliation{State Key Laboratory of Low Dimensional Quantum Physics, Department of Physics, Tsinghua University, Beijing 100084, China}
\affiliation{Collaborative Innovation Center of Quantum Matter, Beijing 100084, China}

\date{\today}

\begin{abstract}
We propose a general scheme for inducing resonant exchange between spins or pseudo-spins
of unmatched levels via periodic driving.
The basic idea is illustrated for a system of two heteronuclear atoms, for which
analytical results are provided for the effective spin exchange (SE) interaction strength. It is then applied to the mixture of
$^{23}$Na and $^{87}$Rb atoms with a radio-frequency (rf) or microwave field near-resonant to the mismatched Zeeman level spacings.
SE interaction engineered this way is applicable to ultracold quantum gas mixtures
involving spinor Bose-Bose, Bose-Fermi, and Fermi-Fermi atoms.
\end{abstract}

\maketitle
\section{Introduction}

Spin exchange (SE) is among the most elementary two body interactions in quantum many body systems.  Between two neutral atoms, this exchange can occur within valence electron spins, nuclear spins, or between the electron and nuclear spins. Its coherent teeterboard-like coupling facilitates excitation exchange between two spinor particles and plays an important role in interesting quantum phenomena ranging from versatile magnetic ordered states such as ferromagnetic or antiferromagnetic phases \cite{Ho1998,Ohmi1998}, collective atomic spin-mixing dynamics in both bosonic~\cite{Pechkis2013,Kuwamoto2004,Schmaljohann2004,Chang2004,Chang2005,Widera2005, Kronjager2006,Black2007,Klempt2009,He2015} and fermionic~\cite{Krauser2012,Krauser2014,PhysRevLett.110.250402,PhysRevA.87.043610} quantum gases, etc. SE can also be employed for spin-squeezing and entangled state generation and preparation in atomic spinor systems \cite{Luo620,Lucke773,Gross2011,PhysRevLett.107.210406}, and for coherence and quantum state transfer in quantum information studies using color centers or NMR techniques \cite{Chen2015,Neumann542,PhysRevLett.102.057403,PhysRevLett.93.130501,Plenio2013,Cai2013}.



SE interaction between heteronuclear atoms is typically small
or even minute in magnitude compared to other energy scales,
such as the density dependent mean field, linear or even quadratic Zeeman shifts, etc.
Controlled SE is thus difficult unless a resonance is encountered.
Between atoms of the same species, this exchange resonance naturally appears
due to their identical pseudo-spin construct, i.e., with the same level spacing,
as has already been studied extensively for spin mixing in $^{87}$Rb atomic Bose-Einstein condensate (BEC) \cite{Kronjager2006,Widera2005}.
If two atoms in the $F=1$ ground states are initially prepared in the $m_F=0$ state,
SE flips one atom spin up into the $m_F=+1$ state while the other one gets flipped down into the $m_F=-1$
, or {\it vice versa}. For $^{87}$Rb atoms, this interaction is calibrated by a spin dependent scattering length $c_2\sim 0.3\,(a_B) <0$,
which denotes a ferromagnetic interaction (with $a_B$ the Bohr radius). It is much smaller than the spin independent
scattering length $c_0\sim 100\,(a_B)>0$. At realized condensate densities, $|c_2|$ is typically not more than a few Hz.
The quadratic Zeeman shift, which differentially detunes the level spacings between
the up ($|m_F=0\rangle\to |m_F=1\rangle$) and down ($|m_F=0\rangle\to |m_F=-1\rangle$) flips,
causes the SE to be off resonant. Thus despite of the null out of the
 linear Zeeman shifts respectively for the up and down spin flips,
observation of coherent spin mixing limits
the background bias $B$ field to be around $1$ Gauss. Further tuning around the resonance
can be accomplished via the ac-stark shifts from a dressing microwave coupled to the $F=2$ manifold \cite{Luo620,Gerbier2006,Zhao2014}. In NMR physics, spin exchange between electronic and nuclear spin can be tunned by Hartmann-Hahn double resonance (HHDR) \cite{HHDR1962,Plenio2013,Cai2013} since nuclear spin is not sensitive to external field.


In addition to spin mixing dynamics,
recent studies in SE also concern the physics associated with interspecies SE interactions in mixtures
of heteronuclear atoms and their properties such as the
ground state phases and entanglement~\cite{Shi2006,Luo2007,Xu2009,Xu2010,Shi2010,Zhang2010,Xu2010b,Xu2011,Shi2011,Xu2012,Li2015}.
The first SE driven coherent heteronuclear spin dynamics are
observed in an ultracold bosonic mixture of ($F=1$) $^{87}$Rb and $^{23}$Na atoms~\cite{Li2015},
which is nicely described by mean field based theories as in single atomic species~\cite{Xu2009,Xu2012}.
The dynamical effort of SE interaction $\propto (s_+^{(a)}s_-^{(b)}+s_-^{(a)}s_+^{(b)})$ between
two unlike ($\eta={a,b}$) spin-$1/2$ atoms ($\vec s^{(\eta)}$) heavily depends on their differential Zeeman shifts.
For the case of $^{87}$Rb and $^{23}$Na atoms in the $F=1$ ground states mentioned above,
their Land{\'e} g-factors are essentially the same because of their equal nuclear and electron spins.
Hence, an accidental interspecies SE resonance occurs at $B_c\sim 1.69\,\rm G$, a small but non-zero $B$ field. More generally, the Land{\'e} g-factors for unlike atoms can be very different, leading to a large Zeeman level spacing mismatch ($\sim$ 1 MHz) even at a moderately low magnetic field ($\sim$ 1 Gauss). Such a large detuning can completely overwhelm the typical rate $|c_2|$ of SE.
The other option of working at a near zero bias $B$ field is difficult due to the 
experimental challenge of controlling the (fluctuating) ambient magnetic field.

\begin{figure}[tbp]
\centering
\includegraphics[width=\linewidth]{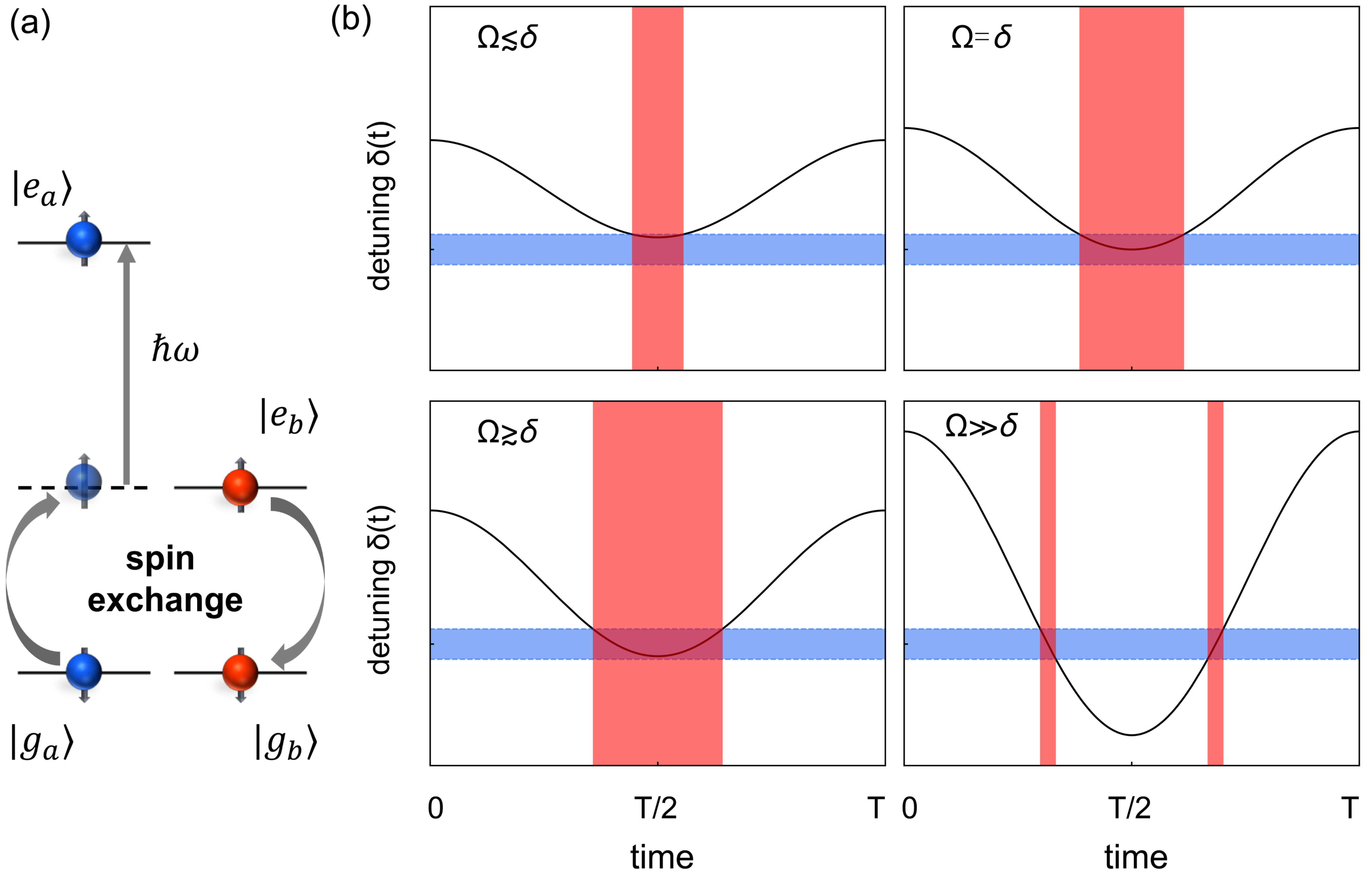}
\caption{(color online). (a) A schematic illustration for interspecies SE assisted by periodic driving.
(b) The time-dependent detuning $\delta(t)$ (black solid line)
in the presence of the drive with a period $T=2\pi/\omega$.
Effective interspecies SE occurs when $|\delta(t)|\le c$ (blue shaded region) for the (red) highlighted time windows
for various driving amplitude $\Omega\lesssim \delta$, $\Omega=\delta$, $\Omega\gtrsim \delta$, and $\Omega\gg \delta$.}
\label{fig1}
\end{figure}

This paper presents a general scheme for promoting resonant SE
between heteronuclear atoms by compensating for their energy level mismatch
using an appropriately modulated $B$ field or rf-field. The basic idea is
illustrated in Fig. \ref{fig1} with the modulation frequency resonant to the
level spacing mismatch. Such a scheme is of course limited to realizable frequency 
ranges of available technologies.
 The different Land{\'e} g-factors for the heteronuclear atoms result in
 the different couplings with the modulated $B$ field.
As we will show in the following tuning the amplitude and/or the frequency of the driving field
controls the interspecies SE dynamics.
We will first illustrate the basic operation of our scheme for a simple model of
two unlike atoms. The result obtained is then applied to a
realistic experiment of $^{87}$Rb and $^{23}$Na mixture,
accompanied with detailed numerical simulations.
Perspective applications to more general cases are then discussed together with
a realistic assessment of the potential restrictions.

\section{Two atom physics}

Without loss of generality,
we assume an isotropic interspecies spin-spin interaction (SSI) of strength $c$ between
the two heteronuclear atoms. The model Hamiltonian thus becomes
\begin{eqnarray}
H &=&\hbar\omega_a s_z^{(a)}+\hbar\omega_b s_z^{(b)}+c\,\bold{s}^{(a)}\cdot\bold{s}^{(b)}+H_D(t), \label{sec1:model_Hamiltonian}\\
H_D(t) &=& \hbar\Omega_a\, s_z^{(a)}\cos{\omega t}+\hbar\Omega_b\, s_z^{(b)}\cos{\omega t},
\label{sec1:effective_Hamiltonian}
\end{eqnarray}
where $s_\mu^{(\eta)}$ ($\mu=x,y,z$, and $\eta=a,b$) denotes the spin-1/2 matrix for atom $\eta$
with level spacing $\hbar\omega_\eta$ between spin up $|e_\eta\rangle$
and down $|g_\eta\rangle$ states.
$H_D(t)$ describes the couplings between atoms and an external periodic driving ($B$) field
along the $z$-axis direction. Other forms of coupling
such as $\propto  s_x^{(\eta)}$ or $\propto s_y^{(\eta)}$ give similar results
and will not be discussed here explicitly.

Even at a small $B$ field, the mismatch between the
pseudo-spin level spacings for two unlike atoms,
can be much larger than their SE interaction, i.e., $\delta=\omega_a-\omega_b\gg |c|/\hbar$, assuming $\omega_a>\omega_b$.
Thus efficient SE dynamics calls for suitable level shifts to compensate for this mismatch.
Ac-stark shift from a microwave field is often employed,
although it provides for only a small $\delta$ \cite{Gerbier2006,Zhao2014}.
Our idea is instead to apply an external $\pi$-polarized oscillating rf or microwave field
with frequency $\omega\sim \delta$.
As illustrated in Fig.~{\ref{fig1}(a), when the above condition is satisfied,
the interspecies SE $|g_a,e_b\rangle \leftrightarrow |e_a,g_b\rangle$
can hit a resonance assisted by the absorption or emission of an oscillation quantum (or photon)
of energy $\hbar\omega$.
The instantaneous level mismatch between the two-atom states $|g_a,e_b\rangle$ and $|e_a,g_b\rangle$
reduces to $\delta(t)=(\omega_a+\Omega_a\cos{\omega t})-(\omega_b+\Omega_b\cos{\omega t})=\delta+\Omega \cos{\omega t}$.
The differential coupling $\Omega\equiv\Omega_a-\Omega_b$ tunes SE into resonance $\delta(t)\sim c$
analogous to differential Zeeman shifts tunes a magnetic Feshbach resonance,
 albeit at selected instants due to the explicit time dependence here.
At a fixed $\omega$, the windows for near-resonant SE within one driving period
are highlighted (red) in Fig.~\ref{fig1}(b) for various driving amplitude.
The largest time window appears for $\Omega\gtrsim \delta$,
which is more rigorously confirmed by the Floquet theory.


\begin{figure}[tbp]
\centering
\includegraphics[width=0.96\linewidth]{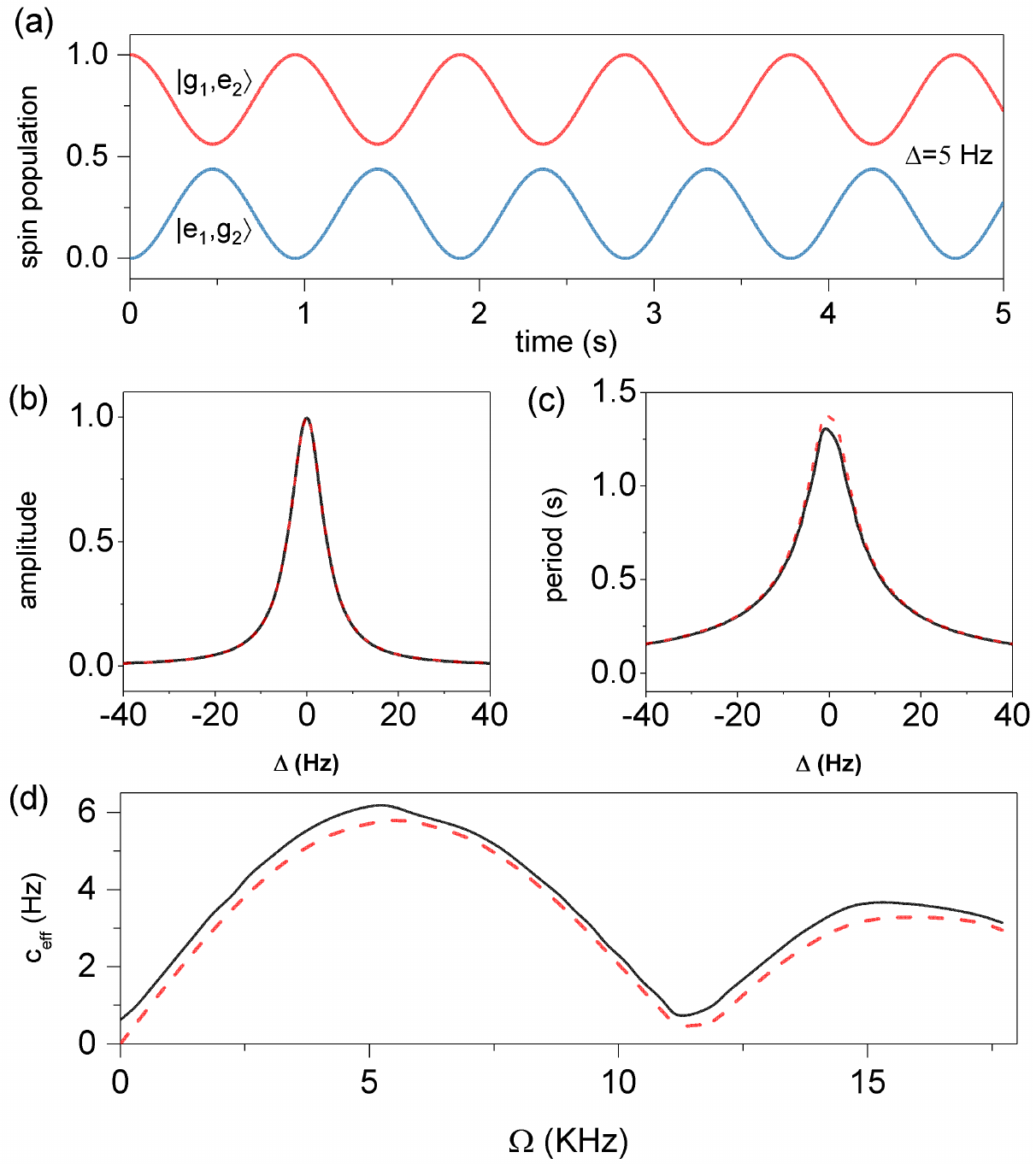}
\caption{(color online). Numerical results compared to analytical ones for $\delta=\omega_a-\omega_b=3\,\rm kHz$ and $c/\hbar=10\,\rm Hz$
with detuning $\Delta=\omega-\delta$. (a) Time evolution of fractional populations for $\Delta=5$ Hz and $\Omega=\omega=\delta$.
Spin oscillation periods (b) and amplitudes (c) from numerical evolutions with the original Hamiltonian Eq.~(\ref{sec1:model_Hamiltonian}) (black solid lines) and the effective Hamiltonian (red dashed lines). (d) The dependence of $c_{\text{eff}}$ on $\Omega$ at $\omega=\delta$.
The red dashed line denotes the analytic formula $c_{\rm eff}=cJ_1(\Omega/\omega)$ while the black solid line is
based on the oscillation periods computed from the dynamics of the original Hamiltonian.}
\label{fig2}
\end{figure}

In the high frequency limit $\omega\sim\delta\gg c/\hbar$,
an effective time-independent Hamiltonian emerges
\begin{eqnarray}
H_{\rm eff}&=& \hbar(\omega_a-{\omega}/{2})s_z^{(a)}+\hbar(\omega_b+{\omega}/{2})s_z^{(b)} \nonumber \\
&&-c_{\text{eff}}\,\bold{s}^{(a)}\cdot\bold{s}^{(b)} +\tilde{c}\,s_z^{(a)}s_z^{(b)},
\label{eqn13}
\end{eqnarray}
as detailed in the appendix below
with $c_{\rm eff}=cJ_1(\Omega/\omega)$ and $\tilde{c}=c[1-J_1(\Omega/\omega)]$.
The minus sign in front of $c_{\text{eff}}$ does not imply that
the SSI has changed its sign entirely due to the follow up term $\propto s_z^{(a)} s_z^{(b)}$.
For our idea to work, the coupling amplitudes for the two atoms
must be different, i.e., $\Omega_a\neq\Omega_b$, or $\Omega\neq 0$ as otherwise $c_{\text{eff}}=0$.
Our proposal thus can be applied when the two atoms are coupled to a driving
field with different strength, a condition that is almost always satisfied for heternuclear
atoms when their pseudo-spin states exhibit different Land{\'e} g-factors.

The analytical results above are confirmed by numerical simulations
for the full dynamics including the periodic drive at
$\delta=\omega_a-\omega_b=3\,\rm kHz$ and $c/\hbar=10\,\rm Hz$ (satisfying $\delta\gg c$).
The simulation starts with the two atoms initially in the state $|g_a,e_b\rangle$.
Figure~\ref{fig2} shows the nice agreement between analytical and numerical results.
The peaks for both the period and amplitude are located at $\Delta=\omega-\delta=0$ as expected.
The numerical result for the effective SE interaction strength, as shown in Figs. \ref{fig2}(d)(red dashed line),
is derived by matching the frequency of spin population oscillation (from Fourier analysis)
to the analytical result $\sqrt{4c_{\text{eff}}^2+\Delta^2}/2$ given by effective Hamiltonian (\ref{eqn13}).
We fix $\Delta=0$ and change $\Omega$ such that $c_{\text{eff}}$
reduces simply to the frequency of spin oscillation.

\section{Spinor mixture of ${}^{87}$$\text{Rb}$ and ${}^{23}$$\text{Na}$}

We next extend the above discussion
for two atoms to a mixture of bosonic spinor $^{23} $Na($\eta=a$) and $^{87}$Rb
($\eta=b$) atoms in the ground $F=1$ states~\cite{Li2015}.
This represents a special case as their level spacing
mismatch is smaller because the nuclear and electronic spins for both atoms are the same.
Their near-resonant interspecies spin dynamics are
recently observed around $B_c\sim 1.6$ (Gauss).
In the off resonant case when their energy level mismatch
is much larger than the interspecies SE strength, this combination still represents a nice
example to test our idea of periodic driving assisted resonant SE.

\begin{figure*}[!htp]
\centering
\includegraphics[scale=0.65]{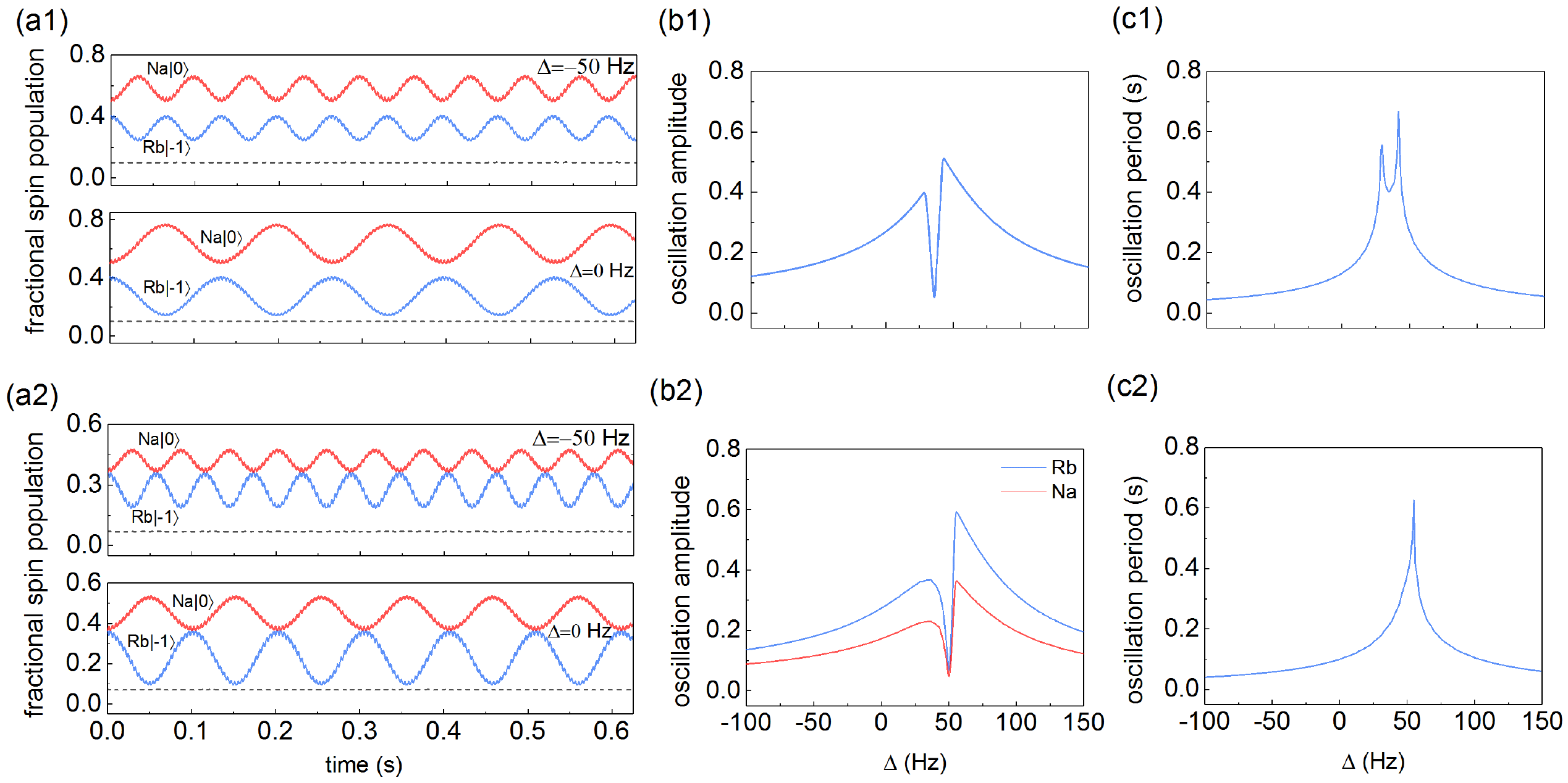}
\caption{The dependence of SE dynamics on $\omega$ for Rb (red line) and Na (blue line) atoms
at $B_0=2.2\,\rm G$ where the Zeeman energy level spacing mismatch between the two spin states $|-1,0\rangle$ and $|0,-1\rangle$
is $\delta\simeq 2\pi\times 227$ Hz and $\Omega=\delta$. (a1-a2) Coherent spin oscillations of balanced (a1) and unbalanced (a2)
atomic populations at different detuning. The black dashed lines denote populations of state $|1\rangle$.
(b1-b2) The dependence of oscillation amplitude on $\Delta$ for balanced (b1) and unbalanced (b2) mixtures.
(c1-c2) The same as above but for the oscillation period in balanced (c1) and unbalanced (c2) mixtures. }
\label{fig3}
\end{figure*}

The model Hamiltonian is detailed in the appendix with
$m_\eta$ the atomic mass, and $\mu=m_1m_2/(m_1+m_2)$
the interspecies reduced mass. $V_\eta$ denotes the trap potential, and
$p_\eta$ and $q_\eta$ are respectively the linear and quadratic Zeeman shifts, while
$c_0^{(\eta)}$ and $c_2^{(\eta)}$ label the intra-atomic density-density and SE interaction strengths.
The interspecies spin-independent, spin-exchange, and spin-singlet pairing interaction strengths
are denoted by $\alpha$, $\beta$, and $\gamma$ as before in studies of binary mixture SE dynamics \cite{Xu2009}
and their values are known to be $(\alpha,\beta,\gamma)=2\pi\hbar^2a_B/\mu\times(78.9,-2.5,0.06)$
for this mixture.

The experiments of Ref. \cite{Li2015} are carried out for a $^{23}$Na atomic BEC with
a cold thermal $^{87}$Rb atomic gas in an optical dipole trap.
Their dynamics are governed by the following coupled equations
\begin{widetext}
\begin{eqnarray}
i\hbar\frac{\partial}{\partial t}\phi&=&
\left[ -\frac{\hbar^2}{2m_a}\nabla^2-p_aF_z +q_aF_z^2+V_a
+c_0^{(a)}\text{Tr}(n_{a})+c_2^{(a)} (\phi^{\dagger}\mathbf{F}\phi)\cdot\mathbf{F} \right]\phi\nonumber\\
&&+[\alpha\text{Tr}(n_b)
+\beta\text{Tr}(\mathbf{F}n_b) \cdot\mathbf{F}+\gamma\mathcal{U}_{b}]\phi,\\
\frac{\partial}{\partial t}f&=&-\frac{\bold{p}}{m_b}\cdot \nabla_{\bold{r}}f+\nabla_{\mathbf{r}}V_b\cdot\nabla_{\mathbf{p}}f
+\frac{1}{i\hbar}[U,f]+\frac{1}{2}\{\nabla_{\mathbf{r}}U,\nabla_{\mathbf{r}}f\},
\end{eqnarray}
with
\begin{eqnarray}
U&=&-p_bF_z+q_bF_{z}^2+c_0^{(b)}\text{Tr}(n_b) +c_0^{(b)}n_b+c_2^{(b)}\text{Tr}(\mathbf{F}n_b)\cdot\mathbf{F}+c_2^{(b)}\mathbf{F}n_b\cdot\mathbf{F}\nonumber\\
&&+\alpha\text{Tr}(n_a)+\beta\text{Tr}(\mathbf{F}n_a)\cdot\mathbf{F}+\gamma\mathcal{U}_a,
\end{eqnarray}
\end{widetext}
where the Na condensate is described by its mean field
$\phi=\langle\hat{\phi}_a \rangle=(\phi_{1},\phi_0,\phi_{-1})^{T}$
 and $(n_{a})_{ij}\equiv\phi^{*}_j\phi_i$,
the Rb gas is described by the collisionless Boltzmann equation
in terms of the Wigner function
$f_{ij}(\bold{r},\bold{p},t)=\langle e^{iHt/\hbar}\hat{f}_{ij}(\bold{r},\bold{p})e^{-iHt/\hbar}\rangle$
and $\hat{f}_{ij}\equiv\int d\bold{r}' e^{-i\bold{p}\cdot\bold{r}/\hbar}\hat{\psi}_j^{\dagger}(\bold{r}-\bold{r}'/2) \hat{\psi}_i(\bold{r}+\bold{r}'/2)$. We define $(n_b(\mathbf{r},t))_{ij}=\int d\mathbf{p} f_{ij}(\mathbf{r},\mathbf{p},t)/(2\pi\hbar)^3$,
$(\mathcal{U}_b)_{ij}=(-1)^{i-j}(n_b)_{\bar{j}\bar{i}}/3$, and
$(\mathcal{U}_a)_{ij}=(-1)^{i-j}(n_a)_{\bar{j}\bar{i}}/3$ with $\bar{i}=-i$.
When one atomic species is non-condensed, the single-mode approximation (SMA)~\cite{Li2015}
is well satisfied for both atomic species. The resulting simplified equations above forms the basis of our numerical study.

The accidental resonance reported in Ref.~\cite{Li2015} at $B_c\sim 1.69\,\rm G$
is between the two atom states
$|m_F^{(a)}=0,m_F^{(b)}=-1\rangle \leftrightarrow |-1,0\rangle$.
Away from this resonance with either increasing or decreasing $B$ field,
the interspecies SE dynamics are suppressed.
Our scheme comes with a $\pi$-polarized periodic rf or microwave field coupled to the atoms
\begin{eqnarray}
H_D(t)=\cos{\omega t}\int d\bold{r}\left\lbrace \hbar\Omega_a\hat{\phi}^{\dagger}F_{z}^{(a)}\hat{\phi} +
\hbar\Omega_b\hat{\psi}^{\dagger}F_{z}^{(b)}\hat{\psi} \right\rbrace . \hskip 12pt
\end{eqnarray}
At $B=2.2\,\rm G$, for instance, the level spacing mismatch between the two atom spin states $|0,-1\rangle$ and $|-1,0\rangle$
 is $\delta\simeq 2\pi\times 227~\rm Hz$, which is much larger than the typical SE strength $\beta$.
The intra-species spin dynamics are also suppressed due to the large quadratic Zeeman shifts at this $B$ field.
We numerically explored this case for both balanced and imbalanced populations of $^{87}$Rb and $^{23}$Na atoms,
starting with a coherent superposition internal states for both species.
To promote strong effective interspecies SE, $\Omega=\delta$ is taken,
and $\omega$ is varied in the vicinity of the two atom resonance $\sim\delta$.
For the balanced case with $N_a=N_b=6\times 10^4$ atoms,
we consider an initial configuration with $50\%$ population of Rb (Na) atoms
in the state $|-1\rangle$ ($|0\rangle$), $40\%$ in $|0\rangle$ ($|-1\rangle$), and $10\%$ in $|+1\rangle$ ($|+1\rangle$).
For the unbalanced case of $N_b=6.33\times 10^4$ and $N_a=10.40\times 10^4$, the initial states
for both atoms are prepared with $36$\% population in state $|0\rangle$, $57\%$ in $|-1\rangle$, and $7\%$ in $|+1\rangle$ approximately.
The resulting near-resonant interspecies SE dynamics are shown in Fig.~\ref{fig3}.
Both the amplitude and period of spin oscillations are found to tune with $\omega$.
The resonance peak is seen to be shifted from the two atom case of $\Delta=0$
due to mean field interactions, while the width of resonance remains of the same order
as that induced by the bare SE interaction strength at weak $B$ field shown in Ref.~\cite{Li2015}.
It is interesting to point out that for the controlled SE dynamics
the periodic external drive introduced
 does not seem to affect other SSI channels
since it does not induce single particle excitation as shown in Figs.~\ref{fig3}(a1,a2) (black dashed line).

Finally we note that our idea for controlled SE as discussed differs from both recently demonstrated
scenario~\cite{Li2015} and the widely known HHDR applied in NV-center \cite{Plenio2013,Cai2013}. 
The first scenario is based on shifting of the resonance field $B_c$
with an optically induced species-dependent (time-independent) static synthetic $B$ field.
Complications to balance the amount of species- and spin-dependent vector light shifts
do not arise in our scheme.
In the second scenario, at least one of the atom system is in strong driving limit and being dressed by the external field. Resonant spin exchange occurs when the dressed states splitting matches the level spacing of another atom. While in our case, the spin state is neither dressed nor flipped by driving filed and collective spin dynamics occurs due to inherent SSI between atoms. Thus our idea is more generally grouped into Floquet engineering and can be applied to tune effective interspecies SE for various types of spinor atomic mixtures.

\section{Conclusion}
In conclusion, we present a general scheme to engineer resonant heteronuclear atomic
spin dynamics by applying a periodic coupling field. This applies for
interatomic species spin dynamics when the Zeeman energy level spacing
mismatch between the two species is much larger than their SSI strength.
Our method is applicable to several ongoing mixture experiments,
and is illustrated for the mixture of $^{23}$Na and $^{87}$Rb atoms
 where spin dynamics were previously observed in the $F=1$ ground stats at near zero field.
A simple calculation using Fermi's golden rule shows that inelastic decay rate
associated with SE collision is about $10^{-14}$ $\text{cm}^{3}\cdot \text{s}^{-1}$
for the $^{23}\text{Na}-^{87}\text{Rb}$ atom mixture, which should provide for a sufficiently long life
time to carry out the proposed periodic modulation experiment.
Another promising candidate system for applying our idea
is the $^6$Li-$^{23}$Na (Fermi-Bose) mixture which exhibits two zero crossings
for the Zeeman level mismatch at $B=0$ G and $B=70.2$ G
between the $|-1/2,1\rangle\leftrightarrow |1/2,0\rangle$ states \cite{ArnoTrautmann2016}.

%

\section*{Acknowledgement}
This work is supported by the National Basic Research Program of China (973 program) (No. 2013CB922004), NSFC (No. 91421305, No. 11574100, No. 11654001, and No. 11374176) and the National Thousand-Young-Talents Program.

\appendix

\section{Derivation of the effective time independent Hamiltonian}

By applying a time-dependent unitary transformation
$U(t)=e^{-i\omega t(s_z^{(a)}-s_z^{(b)})/2-i\int^{t}d\tau ~H_D(\tau)/\hbar}$
for the periodically time-dependent Hamiltonian, we obtain
\begin{eqnarray}
\tilde{H}&=&U^{\dagger}HU-i\hbar U^{\dagger}\partial _tU \nonumber \\
&=&\hbar(\omega_a-\frac{\omega}{2})s_z^{(a)}+\hbar(\omega_b+\frac{\omega}{2})s_z^{(b)}+cA(t)\bold{s}^{(a)}\cdot\bold{s}^{(b)} \nonumber\\
&&+cB(t)(s_x^{(a)}s_y^{(b)}-s_y^{(a)}s_x^{(b)})+cR(t)s_z^{(a)}s_z^{(b)},
\end{eqnarray}
where
\begin{eqnarray}
A(t)&=&\sin{\theta(t)}\sin{\theta'(t)}+\cos{\theta(t)}\cos{\theta'(t)},\\
B(t)&=&\cos{\theta(t)}\sin{\theta'(t)}-\sin{\theta(t)}\cos{\theta'(t)}, \\
R(t)&=&1-A(t), \\
\theta(t) &=& \omega t/2+(\Omega_a/\omega)\sin{\omega t},\\
\theta'(t) &=& -\omega t/2+(\Omega_b/\omega)\sin{\omega t}.
\end{eqnarray}
When $\omega\gg c/\hbar$, the external field oscillates much faster than the internal interspecies spin-exchange dynamics. Thus, the effect of the external driving can be averaged within one oscillating period $T=2\pi/\omega$, i.e. $\langle Q(t) \rangle=(1/T)\int_0^{T}d\tau~Q(\tau)$ with $Q(t)$ being any quantity oscillating with frequency $\omega$. This is equivalent to the high frequency approximation in the general Floquet theory. Based on the identities
\begin{gather}
\langle \cos{(x\sin{\omega t})} \rangle = J_0(x), \\
\langle \cos{(\omega t+x\sin{\omega t})} \rangle = J_1(x), \\
\langle \sin{\theta}\sin{\theta'} \rangle = \frac{1}{2}\left[ J_1(\frac{\Omega}{\omega})-J_0(\frac{\Omega_a+\Omega_b}{\omega})) \right], \\
\langle \sin{\theta}\sin{\theta'} \rangle = -\frac{1}{2}\left[ J_1(\frac{\Omega}{\omega})+J_0(\frac{\Omega_a+\Omega_b}{\omega})) \right], \\
\langle \sin{\theta}\cos{\theta'}\rangle = \langle \cos{\theta}\sin{\theta'} \rangle =0,
\end{gather}
we obtain $\langle A(t) \rangle=J_1(\Omega/\omega)$ and $\langle B(t)\rangle=0$, where $J_0$ and $J_1$ are the Bessel functions of the first kind and $\Omega=\Omega_a-\Omega_b$. Thus, we obtain the effective time-independent Hamiltonian as given in the main text.
\\

\section{A mixture of two spinor atomic gases}

The Hamiltonian for the spin-1 $^{23}$Na (a) and $^{87}$Rb (b) atom mixture as considered in the main text is given by
\begin{eqnarray}
H &=& H_{\text{Rb}}+H_{\text{Na}}+H_{int},\nonumber \\
H_{\text{Na}}&=& \int d\bold{r}~\hat{\phi}_i^{\dagger}\left(-\frac{\hbar^2}{2m_{a}}\nabla^2+V_{a}-p_{a}F_{z}+q_{a}F_{z}^2 \right)\hat{\phi}_i, \nonumber \\
&&+
\frac{c_0^{(a)}}{2}\hat{\phi}_i^{\dagger}\hat{\phi}_j^{\dagger}\hat{\phi}_j\hat{\phi}_i +\frac{c_2^{(a)}}{2}\hat{\phi}_i^{\dagger}\hat{\phi}_k^{\dagger}({\bold{F}})_{ij}\cdot({\bold{F}})_{kl}\hat{\phi}_l\hat{\phi}_j \nonumber\\
H_{\text{Rb}}&=& \int d\bold{r}~\hat{\psi}_i^{\dagger}\left(-\frac{\hbar^2}{2m_{b}}\nabla^2+V_{b}-p_{b}F_z+q_{b}F_z^2 \right)\hat{\psi}_i,\nonumber \\
&&+
\frac{c_0^{(b)}}{2}\hat{\psi}_i^{\dagger}\hat{\psi}_j^{\dagger}\hat{\psi}_j\hat{\psi}_i +\frac{c_2^{(b)}}{2}\hat{\psi}_i^{\dagger}\hat{\psi}_k^{\dagger}({\bold{F}})_{ij}\cdot({\bold{F}})_{kl}\hat{\psi}_l\hat{\psi}_j \nonumber\\
H_{int}&=&\int d\bold{r}~\alpha\hat{\psi}_i^{\dagger}\hat{\phi}_j^{\dagger}\hat{\phi}_j\hat{\psi}_i +\beta\hat{\psi}_i^{\dagger}\hat{\phi}_{k}^{\dagger}({\bold{F}})_{ij}\cdot({\bold{F}})_{kl}\hat{\phi}_l\hat{\psi}_j \nonumber \nonumber\\
&&+
\gamma\frac{(-1)^{i-j}}{3}\hat{\psi}_i^{\dagger}\hat{\phi}_{-i}^{\dagger}\hat{\phi}_{-j}\hat{\psi}_j,
\label{sec3:Hint}
\end{eqnarray}
where $F_{x,y,z}$ are spin-1 matrices, $m_{b},V_{b},p_{b}$ and $q_{b}$ ($m_{a},V_{a},p_{a},q_{a}$) respectively denote the atomic mass, trap potential, linear and quadratic Zeeman shifts of $b$ ($a$) atom. $c_0^{(b)}$ and $c_2^{(b)}$ ($c_0^{(a)},c_2^{(a)}$) are the density-density and the spin-exchange interaction strength between $b$ ($a$) atoms. $\alpha$, $\beta$, and $\gamma$ represent the interspecies spin-independent, spin-exchange, and spin-singlet pairing interaction strength. Their values are known to be $(\alpha,\beta,\gamma)=2\pi\hbar^2a_B/\mu\times(78.9,-2.5,0.06)$, where $\mu$ is the reduced mass of b and a atoms and $a_B$ is the Bohr radius.

In the experiment of Ref. \cite{Li2015}, a spin-1 mixture of a cold thermal $^{87}$Rb gas
 with a $^{23}$Na condensate is prepared in a crossed optical dipole trap.
Their dynamics are governed by the following coupled equations
\begin{widetext}
\begin{eqnarray}
i\hbar\frac{\partial}{\partial t}\phi&=&
\left[ -\frac{\hbar^2}{2m_{a}}\nabla^2-p_{a}F_z +q_{a}F_z^2+V_{a}
+c_0^{(a)}\text{Tr}(n_{a})+c_2^{(a)} (\phi^{\dagger}\mathbf{F}\phi)\cdot\mathbf{F} \right]\phi\nonumber\\
&&+\alpha\text{Tr}(n_{b})\phi
+\beta\text{Tr}(\mathbf{F}n_{b}) \cdot\mathbf{F}\phi+\gamma\mathcal{U}_{\psi}\phi,\\
\frac{\partial}{\partial t}f&=&-\frac{\bold{p}}{m_{b}}\cdot \nabla_{\bold{r}}f+\nabla_{\mathbf{r}}V_{b}\cdot\nabla_{\mathbf{p}}f
+\frac{1}{i\hbar}[U,f]+\frac{1}{2}\{\nabla_{\mathbf{r}}U,\nabla_{\mathbf{r}}f\},
\end{eqnarray}
with
\begin{eqnarray}
U&=&-p_{b}F_z+q_{b}F_{z}^2+c_0^{(b)}\text{Tr}(n_{(b)}) +c_0^{(b)}n_{(b)}+c_2^{(b)}\text{Tr}(\mathbf{F}n_{b})\cdot\mathbf{F}+c_2^{b}\mathbf{F}n_{b}\cdot\mathbf{F}\nonumber\\
&&+\alpha\text{Tr}(n_{a})+\beta\text{Tr}(\mathbf{F}n_{a})\cdot\mathbf{F}+\gamma\mathcal{U}_{\phi},
\end{eqnarray}
\end{widetext}
as given in the main text. In deriving the above equations, we have taken the mean-field approximation for the a condensate with
$\phi=\langle\hat{\phi} \rangle=(\phi_{1},\phi_0,\phi_{-1})^{T}$
 and define $(n_{a})_{ij}=\phi^{*}_j\phi_i$.
The standard collisionless Boltzmann equation is adopted to describe the dynamics of
a thermal b gas with the help of Wigner function
$f_{ij}(\bold{r},\bold{p},t)=\langle e^{iHt/\hbar}\hat{f}_{ij}(\bold{r},\bold{p})e^{-iHt/\hbar}\rangle$, where $\hat{f}_{ij}=\int d\bold{r}' e^{-i\bold{p}\cdot\bold{r}/\hbar}\hat{\psi}_j^{\dagger}(\bold{r}-\bold{r}'/2) \hat{\psi}_i(\bold{r}+\bold{r}'/2)$. We also define $(n_{b}(\mathbf{r},t))_{ij}=\int d\mathbf{p} f_{ij}(\mathbf{r},\mathbf{p},t)/(2\pi\hbar)^3$, $(\mathcal{U}_{\psi})_{ij}=(-1)^{i-j}(n_{b})_{\bar{j}\bar{i}}/3$, and
$(\mathcal{U}_{\phi})_{ij}=(-1)^{i-j}(n_{a})_{\bar{j}\bar{i}}/3$ with $\bar{i}=-i$.

When the confinement trapping is strong~\cite{Li2015}, we can further simplified the above equations
 by adopting the single-mode approximation (SMA) for both atomic species, i.e.,
  take $\phi(\mathbf{r},t)=\tilde{\phi}(\bold{r})\xi(t)$ and $f(\mathbf{r},\mathbf{p},t)=\tilde{f}(\bold{r},\bold{p})\sigma_{ab}(t)$,
where $\tilde{\phi}(\bold{r})$ and $\tilde{f}(\bold{r},\bold{p})$ are the same spatial modes of $^{23}$a and $^{87}$b
spin components and $\text{Tr}\sigma=1$, $\xi^{\dagger}\xi=1$. Defining $\tau_{ij}=\xi_j^*\xi_i$, we thus obtain
\begin{eqnarray}
i\hbar\frac{\partial}{\partial t}\tau&=[U_{\text{BEC}},\tau], \\
i\hbar\frac{\partial}{\partial t}\sigma&=[U_{\text{TG}},\sigma],
\end{eqnarray}
with
\begin{eqnarray}
U_{\text{BEC}}&=&-p_{a}F_z+q_{a}F_z^2 +c_2^{a}\bar{n}^c\text{Tr}(\mathbf{F}\tau)\cdot\mathbf{F}\nonumber \\
&+&\beta\bar{n}^{tc}\sqrt{\frac{N_{b}}{N_{a}}}\text{Tr}(\mathbf{F}\sigma)\cdot\mathbf{F} +\gamma\bar{n}^{tc}\sqrt{\frac{N_{b}}{N_{a}}}\mathcal{U}_{\sigma},\\
U_{\text{TG}}&=&-p_{b}F_z+q_{b}F_z^2 +c_2^{b}\bar{n}^t\text{Tr}(\mathbf{F}\sigma)\cdot\mathbf{F}\nonumber \\
&+&c_2^{b}\bar{n}^t\mathbf{F}\sigma\cdot\mathbf{F} +\beta\bar{n}^{tc}\sqrt{\frac{N_{a}}{N_{b}}}\text{Tr}(\mathbf{F}\tau)\cdot\mathbf{F}\nonumber\\
&+&\gamma\bar{n}^{tc}\sqrt{\frac{N_{a}}{N_{b}}}\mathcal{U}_{\tau},
\end{eqnarray}
where $\bar{n}^{tc}=\int d\bold{r}~[\text{Tr}(n_{b}(\bold{r}))]\text{Tr}[(n_{a}(\bold{r}))]/\sqrt{N_{a}N_{b}}$, $\bar{n}^c=\int d\bold{r}[\text{Tr}(n_{a}(\bold{r}))]^2/N_{a}$, $\bar{n}^t=\int d\bold{r}[\text{Tr}(n_{b}(\bold{r}))]^2/N_{b}$,  $(\mathcal{U}_{\sigma})_{ij}=(-1)^{i-j}\sigma_{\bar{j}\bar{i}}/3$, and $(\mathcal{U}_{\tau})_{ij}=(-1)^{i-j}\tau_{\bar{j}\bar{i}}/3$. $N_{a}$ and $N_{b}$ denote the total numbers of a and b atoms, respectively.

\bibliography{reference}

\begin{thebibliography}{41}%
\makeatletter
\providecommand \@ifxundefined [1]{%
 \@ifx{#1\undefined}
}%
\providecommand \@ifnum [1]{%
 \ifnum #1\expandafter \@firstoftwo
 \else \expandafter \@secondoftwo
 \fi
}%
\providecommand \@ifx [1]{%
 \ifx #1\expandafter \@firstoftwo
 \else \expandafter \@secondoftwo
 \fi
}%
\providecommand \natexlab [1]{#1}%
\providecommand \enquote  [1]{``#1''}%
\providecommand \bibnamefont  [1]{#1}%
\providecommand \bibfnamefont [1]{#1}%
\providecommand \citenamefont [1]{#1}%
\providecommand \href@noop [0]{\@secondoftwo}%
\providecommand \href [0]{\begingroup \@sanitize@url \@href}%
\providecommand \@href[1]{\@@startlink{#1}\@@href}%
\providecommand \@@href[1]{\endgroup#1\@@endlink}%
\providecommand \@sanitize@url [0]{\catcode `\\12\catcode `\$12\catcode
  `\&12\catcode `\#12\catcode `\^12\catcode `\_12\catcode `\%12\relax}%
\providecommand \@@startlink[1]{}%
\providecommand \@@endlink[0]{}%
\providecommand \url  [0]{\begingroup\@sanitize@url \@url }%
\providecommand \@url [1]{\endgroup\@href {#1}{\urlprefix }}%
\providecommand \urlprefix  [0]{URL }%
\providecommand \Eprint [0]{\href }%
\providecommand \doibase [0]{http://dx.doi.org/}%
\providecommand \selectlanguage [0]{\@gobble}%
\providecommand \bibinfo  [0]{\@secondoftwo}%
\providecommand \bibfield  [0]{\@secondoftwo}%
\providecommand \translation [1]{[#1]}%
\providecommand \BibitemOpen [0]{}%
\providecommand \bibitemStop [0]{}%
\providecommand \bibitemNoStop [0]{.\EOS\space}%
\providecommand \EOS [0]{\spacefactor3000\relax}%
\providecommand \BibitemShut  [1]{\csname bibitem#1\endcsname}%
\let\auto@bib@innerbib\@empty
\bibitem [{\citenamefont {Ho}(1998)}]{Ho1998}%
  \BibitemOpen
  \bibfield  {author} {\bibinfo {author} {\bibfnamefont {T.-L.}\ \bibnamefont
  {Ho}},\ }\href {\doibase 10.1103/PhysRevLett.81.742} {\bibfield  {journal}
  {\bibinfo  {journal} {Phys. Rev. Lett.}\ }\textbf {\bibinfo {volume} {81}},\
  \bibinfo {pages} {742} (\bibinfo {year} {1998})}\BibitemShut {NoStop}%
\bibitem [{\citenamefont {Ohmi}\ and\ \citenamefont
  {Machida}(1998)}]{Ohmi1998}%
  \BibitemOpen
  \bibfield  {author} {\bibinfo {author} {\bibfnamefont {T.}~\bibnamefont
  {Ohmi}}\ and\ \bibinfo {author} {\bibfnamefont {K.}~\bibnamefont {Machida}},\
  }\href {\doibase 10.1143/JPSJ.67.1822} {\bibfield  {journal} {\bibinfo
  {journal} {Journal of the Physical Society of Japan}\ }\textbf {\bibinfo
  {volume} {67}},\ \bibinfo {pages} {1822} (\bibinfo {year}
  {1998})}\BibitemShut {NoStop}%
\bibitem [{\citenamefont {Pechkis}\ \emph {et~al.}(2013)\citenamefont
  {Pechkis}, \citenamefont {Wrubel}, \citenamefont {Schwettmann}, \citenamefont
  {Griffin}, \citenamefont {Barnett}, \citenamefont {Tiesinga},\ and\
  \citenamefont {Lett}}]{Pechkis2013}%
  \BibitemOpen
  \bibfield  {author} {\bibinfo {author} {\bibfnamefont {H.~K.}\ \bibnamefont
  {Pechkis}}, \bibinfo {author} {\bibfnamefont {J.~P.}\ \bibnamefont {Wrubel}},
  \bibinfo {author} {\bibfnamefont {A.}~\bibnamefont {Schwettmann}}, \bibinfo
  {author} {\bibfnamefont {P.~F.}\ \bibnamefont {Griffin}}, \bibinfo {author}
  {\bibfnamefont {R.}~\bibnamefont {Barnett}}, \bibinfo {author} {\bibfnamefont
  {E.}~\bibnamefont {Tiesinga}}, \ and\ \bibinfo {author} {\bibfnamefont
  {P.~D.}\ \bibnamefont {Lett}},\ }\href {\doibase
  10.1103/PhysRevLett.111.025301} {\bibfield  {journal} {\bibinfo  {journal}
  {Phys. Rev. Lett.}\ }\textbf {\bibinfo {volume} {111}},\ \bibinfo {pages}
  {025301} (\bibinfo {year} {2013})}\BibitemShut {NoStop}%
\bibitem [{\citenamefont {Kuwamoto}\ \emph {et~al.}(2004)\citenamefont
  {Kuwamoto}, \citenamefont {Araki}, \citenamefont {Eno},\ and\ \citenamefont
  {Hirano}}]{Kuwamoto2004}%
  \BibitemOpen
  \bibfield  {author} {\bibinfo {author} {\bibfnamefont {T.}~\bibnamefont
  {Kuwamoto}}, \bibinfo {author} {\bibfnamefont {K.}~\bibnamefont {Araki}},
  \bibinfo {author} {\bibfnamefont {T.}~\bibnamefont {Eno}}, \ and\ \bibinfo
  {author} {\bibfnamefont {T.}~\bibnamefont {Hirano}},\ }\href {\doibase
  10.1103/PhysRevA.69.063604} {\bibfield  {journal} {\bibinfo  {journal} {Phys.
  Rev. A}\ }\textbf {\bibinfo {volume} {69}},\ \bibinfo {pages} {063604}
  (\bibinfo {year} {2004})}\BibitemShut {NoStop}%
\bibitem [{\citenamefont {Schmaljohann}\ \emph {et~al.}(2004)\citenamefont
  {Schmaljohann}, \citenamefont {Erhard}, \citenamefont {Kronj\"ager},
  \citenamefont {Kottke}, \citenamefont {van Staa}, \citenamefont
  {Cacciapuoti}, \citenamefont {Arlt}, \citenamefont {Bongs},\ and\
  \citenamefont {Sengstock}}]{Schmaljohann2004}%
  \BibitemOpen
  \bibfield  {author} {\bibinfo {author} {\bibfnamefont {H.}~\bibnamefont
  {Schmaljohann}}, \bibinfo {author} {\bibfnamefont {M.}~\bibnamefont
  {Erhard}}, \bibinfo {author} {\bibfnamefont {J.}~\bibnamefont {Kronj\"ager}},
  \bibinfo {author} {\bibfnamefont {M.}~\bibnamefont {Kottke}}, \bibinfo
  {author} {\bibfnamefont {S.}~\bibnamefont {van Staa}}, \bibinfo {author}
  {\bibfnamefont {L.}~\bibnamefont {Cacciapuoti}}, \bibinfo {author}
  {\bibfnamefont {J.~J.}\ \bibnamefont {Arlt}}, \bibinfo {author}
  {\bibfnamefont {K.}~\bibnamefont {Bongs}}, \ and\ \bibinfo {author}
  {\bibfnamefont {K.}~\bibnamefont {Sengstock}},\ }\href {\doibase
  10.1103/PhysRevLett.92.040402} {\bibfield  {journal} {\bibinfo  {journal}
  {Phys. Rev. Lett.}\ }\textbf {\bibinfo {volume} {92}},\ \bibinfo {pages}
  {040402} (\bibinfo {year} {2004})}\BibitemShut {NoStop}%
\bibitem [{\citenamefont {Chang}\ \emph {et~al.}(2004)\citenamefont {Chang},
  \citenamefont {Hamley}, \citenamefont {Barrett}, \citenamefont {Sauer},
  \citenamefont {Fortier}, \citenamefont {Zhang}, \citenamefont {You},\ and\
  \citenamefont {Chapman}}]{Chang2004}%
  \BibitemOpen
  \bibfield  {author} {\bibinfo {author} {\bibfnamefont {M.-S.}\ \bibnamefont
  {Chang}}, \bibinfo {author} {\bibfnamefont {C.~D.}\ \bibnamefont {Hamley}},
  \bibinfo {author} {\bibfnamefont {M.~D.}\ \bibnamefont {Barrett}}, \bibinfo
  {author} {\bibfnamefont {J.~A.}\ \bibnamefont {Sauer}}, \bibinfo {author}
  {\bibfnamefont {K.~M.}\ \bibnamefont {Fortier}}, \bibinfo {author}
  {\bibfnamefont {W.}~\bibnamefont {Zhang}}, \bibinfo {author} {\bibfnamefont
  {L.}~\bibnamefont {You}}, \ and\ \bibinfo {author} {\bibfnamefont {M.~S.}\
  \bibnamefont {Chapman}},\ }\href {\doibase 10.1103/PhysRevLett.92.140403}
  {\bibfield  {journal} {\bibinfo  {journal} {Phys. Rev. Lett.}\ }\textbf
  {\bibinfo {volume} {92}},\ \bibinfo {pages} {140403} (\bibinfo {year}
  {2004})}\BibitemShut {NoStop}%
\bibitem [{\citenamefont {{Chang}}\ \emph {et~al.}(2005)\citenamefont
  {{Chang}}, \citenamefont {{Qin}}, \citenamefont {{Zhang}}, \citenamefont
  {{You}},\ and\ \citenamefont {{Chapman}}}]{Chang2005}%
  \BibitemOpen
  \bibfield  {author} {\bibinfo {author} {\bibfnamefont {M.-S.}\ \bibnamefont
  {{Chang}}}, \bibinfo {author} {\bibfnamefont {Q.}~\bibnamefont {{Qin}}},
  \bibinfo {author} {\bibfnamefont {W.}~\bibnamefont {{Zhang}}}, \bibinfo
  {author} {\bibfnamefont {L.}~\bibnamefont {{You}}}, \ and\ \bibinfo {author}
  {\bibfnamefont {M.~S.}\ \bibnamefont {{Chapman}}},\ }\href {\doibase
  10.1038/nphys153} {\bibfield  {journal} {\bibinfo  {journal} {Nature
  Physics}\ }\textbf {\bibinfo {volume} {1}},\ \bibinfo {pages} {111} (\bibinfo
  {year} {2005})}\BibitemShut {NoStop}%
\bibitem [{\citenamefont {Widera}\ \emph {et~al.}(2005)\citenamefont {Widera},
  \citenamefont {Gerbier}, \citenamefont {F\"olling}, \citenamefont {Gericke},
  \citenamefont {Mandel},\ and\ \citenamefont {Bloch}}]{Widera2005}%
  \BibitemOpen
  \bibfield  {author} {\bibinfo {author} {\bibfnamefont {A.}~\bibnamefont
  {Widera}}, \bibinfo {author} {\bibfnamefont {F.}~\bibnamefont {Gerbier}},
  \bibinfo {author} {\bibfnamefont {S.}~\bibnamefont {F\"olling}}, \bibinfo
  {author} {\bibfnamefont {T.}~\bibnamefont {Gericke}}, \bibinfo {author}
  {\bibfnamefont {O.}~\bibnamefont {Mandel}}, \ and\ \bibinfo {author}
  {\bibfnamefont {I.}~\bibnamefont {Bloch}},\ }\href {\doibase
  10.1103/PhysRevLett.95.190405} {\bibfield  {journal} {\bibinfo  {journal}
  {Phys. Rev. Lett.}\ }\textbf {\bibinfo {volume} {95}},\ \bibinfo {pages}
  {190405} (\bibinfo {year} {2005})}\BibitemShut {NoStop}%
\bibitem [{\citenamefont {Kronj\"ager}\ \emph {et~al.}(2006)\citenamefont
  {Kronj\"ager}, \citenamefont {Becker}, \citenamefont {Navez}, \citenamefont
  {Bongs},\ and\ \citenamefont {Sengstock}}]{Kronjager2006}%
  \BibitemOpen
  \bibfield  {author} {\bibinfo {author} {\bibfnamefont {J.}~\bibnamefont
  {Kronj\"ager}}, \bibinfo {author} {\bibfnamefont {C.}~\bibnamefont {Becker}},
  \bibinfo {author} {\bibfnamefont {P.}~\bibnamefont {Navez}}, \bibinfo
  {author} {\bibfnamefont {K.}~\bibnamefont {Bongs}}, \ and\ \bibinfo {author}
  {\bibfnamefont {K.}~\bibnamefont {Sengstock}},\ }\href {\doibase
  10.1103/PhysRevLett.97.110404} {\bibfield  {journal} {\bibinfo  {journal}
  {Phys. Rev. Lett.}\ }\textbf {\bibinfo {volume} {97}},\ \bibinfo {pages}
  {110404} (\bibinfo {year} {2006})}\BibitemShut {NoStop}%
\bibitem [{\citenamefont {Black}\ \emph {et~al.}(2007)\citenamefont {Black},
  \citenamefont {Gomez}, \citenamefont {Turner}, \citenamefont {Jung},\ and\
  \citenamefont {Lett}}]{Black2007}%
  \BibitemOpen
  \bibfield  {author} {\bibinfo {author} {\bibfnamefont {A.~T.}\ \bibnamefont
  {Black}}, \bibinfo {author} {\bibfnamefont {E.}~\bibnamefont {Gomez}},
  \bibinfo {author} {\bibfnamefont {L.~D.}\ \bibnamefont {Turner}}, \bibinfo
  {author} {\bibfnamefont {S.}~\bibnamefont {Jung}}, \ and\ \bibinfo {author}
  {\bibfnamefont {P.~D.}\ \bibnamefont {Lett}},\ }\href {\doibase
  10.1103/PhysRevLett.99.070403} {\bibfield  {journal} {\bibinfo  {journal}
  {Phys. Rev. Lett.}\ }\textbf {\bibinfo {volume} {99}},\ \bibinfo {pages}
  {070403} (\bibinfo {year} {2007})}\BibitemShut {NoStop}%
\bibitem [{\citenamefont {Klempt}\ \emph {et~al.}(2009)\citenamefont {Klempt},
  \citenamefont {Topic}, \citenamefont {Gebreyesus}, \citenamefont {Scherer},
  \citenamefont {Henninger}, \citenamefont {Hyllus}, \citenamefont {Ertmer},
  \citenamefont {Santos},\ and\ \citenamefont {Arlt}}]{Klempt2009}%
  \BibitemOpen
  \bibfield  {author} {\bibinfo {author} {\bibfnamefont {C.}~\bibnamefont
  {Klempt}}, \bibinfo {author} {\bibfnamefont {O.}~\bibnamefont {Topic}},
  \bibinfo {author} {\bibfnamefont {G.}~\bibnamefont {Gebreyesus}}, \bibinfo
  {author} {\bibfnamefont {M.}~\bibnamefont {Scherer}}, \bibinfo {author}
  {\bibfnamefont {T.}~\bibnamefont {Henninger}}, \bibinfo {author}
  {\bibfnamefont {P.}~\bibnamefont {Hyllus}}, \bibinfo {author} {\bibfnamefont
  {W.}~\bibnamefont {Ertmer}}, \bibinfo {author} {\bibfnamefont
  {L.}~\bibnamefont {Santos}}, \ and\ \bibinfo {author} {\bibfnamefont {J.~J.}\
  \bibnamefont {Arlt}},\ }\href {\doibase 10.1103/PhysRevLett.103.195302}
  {\bibfield  {journal} {\bibinfo  {journal} {Phys. Rev. Lett.}\ }\textbf
  {\bibinfo {volume} {103}},\ \bibinfo {pages} {195302} (\bibinfo {year}
  {2009})}\BibitemShut {NoStop}%
\bibitem [{\citenamefont {He}\ \emph {et~al.}(2015)\citenamefont {He},
  \citenamefont {Zhu}, \citenamefont {Li}, \citenamefont {Wang}, \citenamefont
  {Xu},\ and\ \citenamefont {Wang}}]{He2015}%
  \BibitemOpen
  \bibfield  {author} {\bibinfo {author} {\bibfnamefont {X.}~\bibnamefont
  {He}}, \bibinfo {author} {\bibfnamefont {B.}~\bibnamefont {Zhu}}, \bibinfo
  {author} {\bibfnamefont {X.}~\bibnamefont {Li}}, \bibinfo {author}
  {\bibfnamefont {F.}~\bibnamefont {Wang}}, \bibinfo {author} {\bibfnamefont
  {Z.-F.}\ \bibnamefont {Xu}}, \ and\ \bibinfo {author} {\bibfnamefont
  {D.}~\bibnamefont {Wang}},\ }\href {\doibase 10.1103/PhysRevA.91.033635}
  {\bibfield  {journal} {\bibinfo  {journal} {Phys. Rev. A}\ }\textbf {\bibinfo
  {volume} {91}},\ \bibinfo {pages} {033635} (\bibinfo {year}
  {2015})}\BibitemShut {NoStop}%
\bibitem [{\citenamefont {{Krauser}}\ \emph {et~al.}(2012)\citenamefont
  {{Krauser}}, \citenamefont {{Heinze}}, \citenamefont {{Fl{\"a}schner}},
  \citenamefont {{G{\"o}tze}}, \citenamefont {{J{\"u}rgensen}}, \citenamefont
  {{L{\"u}hmann}}, \citenamefont {{Becker}},\ and\ \citenamefont
  {{Sengstock}}}]{Krauser2012}%
  \BibitemOpen
  \bibfield  {author} {\bibinfo {author} {\bibfnamefont {J.~S.}\ \bibnamefont
  {{Krauser}}}, \bibinfo {author} {\bibfnamefont {J.}~\bibnamefont {{Heinze}}},
  \bibinfo {author} {\bibfnamefont {N.}~\bibnamefont {{Fl{\"a}schner}}},
  \bibinfo {author} {\bibfnamefont {S.}~\bibnamefont {{G{\"o}tze}}}, \bibinfo
  {author} {\bibfnamefont {O.}~\bibnamefont {{J{\"u}rgensen}}}, \bibinfo
  {author} {\bibfnamefont {D.-S.}\ \bibnamefont {{L{\"u}hmann}}}, \bibinfo
  {author} {\bibfnamefont {C.}~\bibnamefont {{Becker}}}, \ and\ \bibinfo
  {author} {\bibfnamefont {K.}~\bibnamefont {{Sengstock}}},\ }\href {\doibase
  10.1038/nphys2409} {\bibfield  {journal} {\bibinfo  {journal} {Nature
  Physics}\ }\textbf {\bibinfo {volume} {8}},\ \bibinfo {pages} {813} (\bibinfo
  {year} {2012})}\BibitemShut {NoStop}%
\bibitem [{\citenamefont {{Krauser}}\ \emph {et~al.}(2014)\citenamefont
  {{Krauser}}, \citenamefont {{Ebling}}, \citenamefont {{Fl{\"a}schner}},
  \citenamefont {{Heinze}}, \citenamefont {{Sengstock}}, \citenamefont
  {{Lewenstein}}, \citenamefont {{Eckardt}},\ and\ \citenamefont
  {{Becker}}}]{Krauser2014}%
  \BibitemOpen
  \bibfield  {author} {\bibinfo {author} {\bibfnamefont {J.~S.}\ \bibnamefont
  {{Krauser}}}, \bibinfo {author} {\bibfnamefont {U.}~\bibnamefont {{Ebling}}},
  \bibinfo {author} {\bibfnamefont {N.}~\bibnamefont {{Fl{\"a}schner}}},
  \bibinfo {author} {\bibfnamefont {J.}~\bibnamefont {{Heinze}}}, \bibinfo
  {author} {\bibfnamefont {K.}~\bibnamefont {{Sengstock}}}, \bibinfo {author}
  {\bibfnamefont {M.}~\bibnamefont {{Lewenstein}}}, \bibinfo {author}
  {\bibfnamefont {A.}~\bibnamefont {{Eckardt}}}, \ and\ \bibinfo {author}
  {\bibfnamefont {C.}~\bibnamefont {{Becker}}},\ }\href {\doibase
  10.1126/science.1244059} {\bibfield  {journal} {\bibinfo  {journal}
  {Science}\ }\textbf {\bibinfo {volume} {343}},\ \bibinfo {pages} {157}
  (\bibinfo {year} {2014})}\BibitemShut {NoStop}%
\bibitem [{\citenamefont {Heinze}\ \emph {et~al.}(2013)\citenamefont {Heinze},
  \citenamefont {Krauser}, \citenamefont {Fl\"aschner}, \citenamefont
  {Sengstock}, \citenamefont {Becker}, \citenamefont {Ebling}, \citenamefont
  {Eckardt},\ and\ \citenamefont {Lewenstein}}]{PhysRevLett.110.250402}%
  \BibitemOpen
  \bibfield  {author} {\bibinfo {author} {\bibfnamefont {J.}~\bibnamefont
  {Heinze}}, \bibinfo {author} {\bibfnamefont {J.~S.}\ \bibnamefont {Krauser}},
  \bibinfo {author} {\bibfnamefont {N.}~\bibnamefont {Fl\"aschner}}, \bibinfo
  {author} {\bibfnamefont {K.}~\bibnamefont {Sengstock}}, \bibinfo {author}
  {\bibfnamefont {C.}~\bibnamefont {Becker}}, \bibinfo {author} {\bibfnamefont
  {U.}~\bibnamefont {Ebling}}, \bibinfo {author} {\bibfnamefont
  {A.}~\bibnamefont {Eckardt}}, \ and\ \bibinfo {author} {\bibfnamefont
  {M.}~\bibnamefont {Lewenstein}},\ }\href {\doibase
  10.1103/PhysRevLett.110.250402} {\bibfield  {journal} {\bibinfo  {journal}
  {Phys. Rev. Lett.}\ }\textbf {\bibinfo {volume} {110}},\ \bibinfo {pages}
  {250402} (\bibinfo {year} {2013})}\BibitemShut {NoStop}%
\bibitem [{\citenamefont {Dong}\ and\ \citenamefont
  {Pu}(2013)}]{PhysRevA.87.043610}%
  \BibitemOpen
  \bibfield  {author} {\bibinfo {author} {\bibfnamefont {Y.}~\bibnamefont
  {Dong}}\ and\ \bibinfo {author} {\bibfnamefont {H.}~\bibnamefont {Pu}},\
  }\href {\doibase 10.1103/PhysRevA.87.043610} {\bibfield  {journal} {\bibinfo
  {journal} {Phys. Rev. A}\ }\textbf {\bibinfo {volume} {87}},\ \bibinfo
  {pages} {043610} (\bibinfo {year} {2013})}\BibitemShut {NoStop}%
\bibitem [{\citenamefont {Luo}\ \emph {et~al.}(2017)\citenamefont {Luo},
  \citenamefont {Zou}, \citenamefont {Wu}, \citenamefont {Liu}, \citenamefont
  {Han}, \citenamefont {Tey},\ and\ \citenamefont {You}}]{Luo620}%
  \BibitemOpen
  \bibfield  {author} {\bibinfo {author} {\bibfnamefont {X.-Y.}\ \bibnamefont
  {Luo}}, \bibinfo {author} {\bibfnamefont {Y.-Q.}\ \bibnamefont {Zou}},
  \bibinfo {author} {\bibfnamefont {L.-N.}\ \bibnamefont {Wu}}, \bibinfo
  {author} {\bibfnamefont {Q.}~\bibnamefont {Liu}}, \bibinfo {author}
  {\bibfnamefont {M.-F.}\ \bibnamefont {Han}}, \bibinfo {author} {\bibfnamefont
  {M.~K.}\ \bibnamefont {Tey}}, \ and\ \bibinfo {author} {\bibfnamefont
  {L.}~\bibnamefont {You}},\ }\href {\doibase 10.1126/science.aag1106}
  {\bibfield  {journal} {\bibinfo  {journal} {Science}\ }\textbf {\bibinfo
  {volume} {355}},\ \bibinfo {pages} {620} (\bibinfo {year}
  {2017})}\BibitemShut {NoStop}%
\bibitem [{\citenamefont {L{\"u}cke}\ \emph {et~al.}(2011)\citenamefont
  {L{\"u}cke}, \citenamefont {Scherer}, \citenamefont {Kruse}, \citenamefont
  {Pezz{\'e}}, \citenamefont {Deuretzbacher}, \citenamefont {Hyllus},
  \citenamefont {Topic}, \citenamefont {Peise}, \citenamefont {Ertmer},
  \citenamefont {Arlt}, \citenamefont {Santos}, \citenamefont {Smerzi},\ and\
  \citenamefont {Klempt}}]{Lucke773}%
  \BibitemOpen
  \bibfield  {author} {\bibinfo {author} {\bibfnamefont {B.}~\bibnamefont
  {L{\"u}cke}}, \bibinfo {author} {\bibfnamefont {M.}~\bibnamefont {Scherer}},
  \bibinfo {author} {\bibfnamefont {J.}~\bibnamefont {Kruse}}, \bibinfo
  {author} {\bibfnamefont {L.}~\bibnamefont {Pezz{\'e}}}, \bibinfo {author}
  {\bibfnamefont {F.}~\bibnamefont {Deuretzbacher}}, \bibinfo {author}
  {\bibfnamefont {P.}~\bibnamefont {Hyllus}}, \bibinfo {author} {\bibfnamefont
  {O.}~\bibnamefont {Topic}}, \bibinfo {author} {\bibfnamefont
  {J.}~\bibnamefont {Peise}}, \bibinfo {author} {\bibfnamefont
  {W.}~\bibnamefont {Ertmer}}, \bibinfo {author} {\bibfnamefont
  {J.}~\bibnamefont {Arlt}}, \bibinfo {author} {\bibfnamefont {L.}~\bibnamefont
  {Santos}}, \bibinfo {author} {\bibfnamefont {A.}~\bibnamefont {Smerzi}}, \
  and\ \bibinfo {author} {\bibfnamefont {C.}~\bibnamefont {Klempt}},\ }\href
  {\doibase 10.1126/science.1208798} {\bibfield  {journal} {\bibinfo  {journal}
  {Science}\ }\textbf {\bibinfo {volume} {334}},\ \bibinfo {pages} {773}
  (\bibinfo {year} {2011})}\BibitemShut {NoStop}%
\bibitem [{\citenamefont {Gross}\ \emph {et~al.}(2011)\citenamefont {Gross},
  \citenamefont {Strobel}, \citenamefont {Nicklas}, \citenamefont {Zibold},
  \citenamefont {Bar-Gill}, \citenamefont {Kurizki},\ and\ \citenamefont
  {Oberthaler}}]{Gross2011}%
  \BibitemOpen
  \bibfield  {author} {\bibinfo {author} {\bibfnamefont {C.}~\bibnamefont
  {Gross}}, \bibinfo {author} {\bibfnamefont {H.}~\bibnamefont {Strobel}},
  \bibinfo {author} {\bibfnamefont {E.}~\bibnamefont {Nicklas}}, \bibinfo
  {author} {\bibfnamefont {T.}~\bibnamefont {Zibold}}, \bibinfo {author}
  {\bibfnamefont {N.}~\bibnamefont {Bar-Gill}}, \bibinfo {author}
  {\bibfnamefont {G.}~\bibnamefont {Kurizki}}, \ and\ \bibinfo {author}
  {\bibfnamefont {M.~K.}\ \bibnamefont {Oberthaler}},\ }\href {\doibase
  10.1038/nature10654} {\bibfield  {journal} {\bibinfo  {journal} {Nature}\
  }\textbf {\bibinfo {volume} {480}},\ \bibinfo {pages} {219} (\bibinfo {year}
  {2011})}\BibitemShut {NoStop}%
\bibitem [{\citenamefont {Bookjans}\ \emph {et~al.}(2011)\citenamefont
  {Bookjans}, \citenamefont {Hamley},\ and\ \citenamefont
  {Chapman}}]{PhysRevLett.107.210406}%
  \BibitemOpen
  \bibfield  {author} {\bibinfo {author} {\bibfnamefont {E.~M.}\ \bibnamefont
  {Bookjans}}, \bibinfo {author} {\bibfnamefont {C.~D.}\ \bibnamefont
  {Hamley}}, \ and\ \bibinfo {author} {\bibfnamefont {M.~S.}\ \bibnamefont
  {Chapman}},\ }\href {\doibase 10.1103/PhysRevLett.107.210406} {\bibfield
  {journal} {\bibinfo  {journal} {Phys. Rev. Lett.}\ }\textbf {\bibinfo
  {volume} {107}},\ \bibinfo {pages} {210406} (\bibinfo {year}
  {2011})}\BibitemShut {NoStop}%
\bibitem [{\citenamefont {Chen}\ \emph {et~al.}(2015)\citenamefont {Chen},
  \citenamefont {Schwarz}, \citenamefont {Jelezko}, \citenamefont {Retzker},\
  and\ \citenamefont {Plenio}}]{Chen2015}%
  \BibitemOpen
  \bibfield  {author} {\bibinfo {author} {\bibfnamefont {Q.}~\bibnamefont
  {Chen}}, \bibinfo {author} {\bibfnamefont {I.}~\bibnamefont {Schwarz}},
  \bibinfo {author} {\bibfnamefont {F.}~\bibnamefont {Jelezko}}, \bibinfo
  {author} {\bibfnamefont {A.}~\bibnamefont {Retzker}}, \ and\ \bibinfo
  {author} {\bibfnamefont {M.~B.}\ \bibnamefont {Plenio}},\ }\href {\doibase
  10.1103/PhysRevB.92.184420} {\bibfield  {journal} {\bibinfo  {journal}
  {Physical Review B}\ }\textbf {\bibinfo {volume} {92}},\ \bibinfo {pages}
  {184420} (\bibinfo {year} {2015})}\BibitemShut {NoStop}%
\bibitem [{\citenamefont {Neumann}\ \emph {et~al.}(2010)\citenamefont
  {Neumann}, \citenamefont {Beck}, \citenamefont {Steiner}, \citenamefont
  {Rempp}, \citenamefont {Fedder}, \citenamefont {Hemmer}, \citenamefont
  {Wrachtrup},\ and\ \citenamefont {Jelezko}}]{Neumann542}%
  \BibitemOpen
  \bibfield  {author} {\bibinfo {author} {\bibfnamefont {P.}~\bibnamefont
  {Neumann}}, \bibinfo {author} {\bibfnamefont {J.}~\bibnamefont {Beck}},
  \bibinfo {author} {\bibfnamefont {M.}~\bibnamefont {Steiner}}, \bibinfo
  {author} {\bibfnamefont {F.}~\bibnamefont {Rempp}}, \bibinfo {author}
  {\bibfnamefont {H.}~\bibnamefont {Fedder}}, \bibinfo {author} {\bibfnamefont
  {P.~R.}\ \bibnamefont {Hemmer}}, \bibinfo {author} {\bibfnamefont
  {J.}~\bibnamefont {Wrachtrup}}, \ and\ \bibinfo {author} {\bibfnamefont
  {F.}~\bibnamefont {Jelezko}},\ }\href {\doibase 10.1126/science.1189075}
  {\bibfield  {journal} {\bibinfo  {journal} {Science}\ }\textbf {\bibinfo
  {volume} {329}},\ \bibinfo {pages} {542} (\bibinfo {year}
  {2010})}\BibitemShut {NoStop}%
\bibitem [{\citenamefont {Jacques}\ \emph {et~al.}(2009)\citenamefont
  {Jacques}, \citenamefont {Neumann}, \citenamefont {Beck}, \citenamefont
  {Markham}, \citenamefont {Twitchen}, \citenamefont {Meijer}, \citenamefont
  {Kaiser}, \citenamefont {Balasubramanian}, \citenamefont {Jelezko},\ and\
  \citenamefont {Wrachtrup}}]{PhysRevLett.102.057403}%
  \BibitemOpen
  \bibfield  {author} {\bibinfo {author} {\bibfnamefont {V.}~\bibnamefont
  {Jacques}}, \bibinfo {author} {\bibfnamefont {P.}~\bibnamefont {Neumann}},
  \bibinfo {author} {\bibfnamefont {J.}~\bibnamefont {Beck}}, \bibinfo {author}
  {\bibfnamefont {M.}~\bibnamefont {Markham}}, \bibinfo {author} {\bibfnamefont
  {D.}~\bibnamefont {Twitchen}}, \bibinfo {author} {\bibfnamefont
  {J.}~\bibnamefont {Meijer}}, \bibinfo {author} {\bibfnamefont
  {F.}~\bibnamefont {Kaiser}}, \bibinfo {author} {\bibfnamefont
  {G.}~\bibnamefont {Balasubramanian}}, \bibinfo {author} {\bibfnamefont
  {F.}~\bibnamefont {Jelezko}}, \ and\ \bibinfo {author} {\bibfnamefont
  {J.}~\bibnamefont {Wrachtrup}},\ }\href {\doibase
  10.1103/PhysRevLett.102.057403} {\bibfield  {journal} {\bibinfo  {journal}
  {Phys. Rev. Lett.}\ }\textbf {\bibinfo {volume} {102}},\ \bibinfo {pages}
  {057403} (\bibinfo {year} {2009})}\BibitemShut {NoStop}%
\bibitem [{\citenamefont {Jelezko}\ \emph {et~al.}(2004)\citenamefont
  {Jelezko}, \citenamefont {Gaebel}, \citenamefont {Popa}, \citenamefont
  {Domhan}, \citenamefont {Gruber},\ and\ \citenamefont
  {Wrachtrup}}]{PhysRevLett.93.130501}%
  \BibitemOpen
  \bibfield  {author} {\bibinfo {author} {\bibfnamefont {F.}~\bibnamefont
  {Jelezko}}, \bibinfo {author} {\bibfnamefont {T.}~\bibnamefont {Gaebel}},
  \bibinfo {author} {\bibfnamefont {I.}~\bibnamefont {Popa}}, \bibinfo {author}
  {\bibfnamefont {M.}~\bibnamefont {Domhan}}, \bibinfo {author} {\bibfnamefont
  {A.}~\bibnamefont {Gruber}}, \ and\ \bibinfo {author} {\bibfnamefont
  {J.}~\bibnamefont {Wrachtrup}},\ }\href {\doibase
  10.1103/PhysRevLett.93.130501} {\bibfield  {journal} {\bibinfo  {journal}
  {Phys. Rev. Lett.}\ }\textbf {\bibinfo {volume} {93}},\ \bibinfo {pages}
  {130501} (\bibinfo {year} {2004})}\BibitemShut {NoStop}%
\bibitem [{\citenamefont {London}\ \emph {et~al.}(2013)\citenamefont {London},
  \citenamefont {Scheuer}, \citenamefont {Cai}, \citenamefont {Schwarz},
  \citenamefont {Retzker}, \citenamefont {Plenio}, \citenamefont {Katagiri},
  \citenamefont {Teraji}, \citenamefont {Koizumi}, \citenamefont {Isoya},
  \citenamefont {Fischer}, \citenamefont {McGuinness}, \citenamefont
  {Naydenov},\ and\ \citenamefont {Jelezko}}]{Plenio2013}%
  \BibitemOpen
  \bibfield  {author} {\bibinfo {author} {\bibfnamefont {P.}~\bibnamefont
  {London}}, \bibinfo {author} {\bibfnamefont {J.}~\bibnamefont {Scheuer}},
  \bibinfo {author} {\bibfnamefont {J.-M.}\ \bibnamefont {Cai}}, \bibinfo
  {author} {\bibfnamefont {I.}~\bibnamefont {Schwarz}}, \bibinfo {author}
  {\bibfnamefont {A.}~\bibnamefont {Retzker}}, \bibinfo {author} {\bibfnamefont
  {M.~B.}\ \bibnamefont {Plenio}}, \bibinfo {author} {\bibfnamefont
  {M.}~\bibnamefont {Katagiri}}, \bibinfo {author} {\bibfnamefont
  {T.}~\bibnamefont {Teraji}}, \bibinfo {author} {\bibfnamefont
  {S.}~\bibnamefont {Koizumi}}, \bibinfo {author} {\bibfnamefont
  {J.}~\bibnamefont {Isoya}}, \bibinfo {author} {\bibfnamefont
  {R.}~\bibnamefont {Fischer}}, \bibinfo {author} {\bibfnamefont {L.~P.}\
  \bibnamefont {McGuinness}}, \bibinfo {author} {\bibfnamefont
  {B.}~\bibnamefont {Naydenov}}, \ and\ \bibinfo {author} {\bibfnamefont
  {F.}~\bibnamefont {Jelezko}},\ }\href {\doibase
  10.1103/PhysRevLett.111.067601} {\bibfield  {journal} {\bibinfo  {journal}
  {Phys. Rev. Lett.}\ }\textbf {\bibinfo {volume} {111}},\ \bibinfo {pages}
  {067601} (\bibinfo {year} {2013})}\BibitemShut {NoStop}%
\bibitem [{\citenamefont {Cai}\ \emph {et~al.}(2013)\citenamefont {Cai},
  \citenamefont {Jelezko}, \citenamefont {Plenio},\ and\ \citenamefont
  {Retzker}}]{Cai2013}%
  \BibitemOpen
  \bibfield  {author} {\bibinfo {author} {\bibfnamefont {J.}~\bibnamefont
  {Cai}}, \bibinfo {author} {\bibfnamefont {F.}~\bibnamefont {Jelezko}},
  \bibinfo {author} {\bibfnamefont {M.~B.}\ \bibnamefont {Plenio}}, \ and\
  \bibinfo {author} {\bibfnamefont {A.}~\bibnamefont {Retzker}},\ }\href
  {http://stacks.iop.org/1367-2630/15/i=1/a=013020} {\bibfield  {journal}
  {\bibinfo  {journal} {New Journal of Physics}\ }\textbf {\bibinfo {volume}
  {15}},\ \bibinfo {pages} {013020} (\bibinfo {year} {2013})}\BibitemShut
  {NoStop}%
\bibitem [{\citenamefont {Gerbier}\ \emph {et~al.}(2006)\citenamefont
  {Gerbier}, \citenamefont {Widera}, \citenamefont {F\"olling}, \citenamefont
  {Mandel},\ and\ \citenamefont {Bloch}}]{Gerbier2006}%
  \BibitemOpen
  \bibfield  {author} {\bibinfo {author} {\bibfnamefont {F.}~\bibnamefont
  {Gerbier}}, \bibinfo {author} {\bibfnamefont {A.}~\bibnamefont {Widera}},
  \bibinfo {author} {\bibfnamefont {S.}~\bibnamefont {F\"olling}}, \bibinfo
  {author} {\bibfnamefont {O.}~\bibnamefont {Mandel}}, \ and\ \bibinfo {author}
  {\bibfnamefont {I.}~\bibnamefont {Bloch}},\ }\href {\doibase
  10.1103/PhysRevA.73.041602} {\bibfield  {journal} {\bibinfo  {journal} {Phys.
  Rev. A}\ }\textbf {\bibinfo {volume} {73}},\ \bibinfo {pages} {041602}
  (\bibinfo {year} {2006})}\BibitemShut {NoStop}%
\bibitem [{\citenamefont {Zhao}\ \emph {et~al.}(2014)\citenamefont {Zhao},
  \citenamefont {Jiang}, \citenamefont {Tang}, \citenamefont {Webb},\ and\
  \citenamefont {Liu}}]{Zhao2014}%
  \BibitemOpen
  \bibfield  {author} {\bibinfo {author} {\bibfnamefont {L.}~\bibnamefont
  {Zhao}}, \bibinfo {author} {\bibfnamefont {J.}~\bibnamefont {Jiang}},
  \bibinfo {author} {\bibfnamefont {T.}~\bibnamefont {Tang}}, \bibinfo {author}
  {\bibfnamefont {M.}~\bibnamefont {Webb}}, \ and\ \bibinfo {author}
  {\bibfnamefont {Y.}~\bibnamefont {Liu}},\ }\href {\doibase
  10.1103/PhysRevA.89.023608} {\bibfield  {journal} {\bibinfo  {journal} {Phys.
  Rev. A}\ }\textbf {\bibinfo {volume} {89}},\ \bibinfo {pages} {023608}
  (\bibinfo {year} {2014})}\BibitemShut {NoStop}%
\bibitem [{\citenamefont {Hartmann}\ and\ \citenamefont
  {Hahn}(1962)}]{HHDR1962}%
  \BibitemOpen
  \bibfield  {author} {\bibinfo {author} {\bibfnamefont {S.~R.}\ \bibnamefont
  {Hartmann}}\ and\ \bibinfo {author} {\bibfnamefont {E.~L.}\ \bibnamefont
  {Hahn}},\ }\href {\doibase 10.1103/PhysRev.128.2042} {\bibfield  {journal}
  {\bibinfo  {journal} {Phys. Rev.}\ }\textbf {\bibinfo {volume} {128}},\
  \bibinfo {pages} {2042} (\bibinfo {year} {1962})}\BibitemShut {NoStop}%
\bibitem [{\citenamefont {Shi}\ and\ \citenamefont {Niu}(2006)}]{Shi2006}%
  \BibitemOpen
  \bibfield  {author} {\bibinfo {author} {\bibfnamefont {Y.}~\bibnamefont
  {Shi}}\ and\ \bibinfo {author} {\bibfnamefont {Q.}~\bibnamefont {Niu}},\
  }\href {\doibase 10.1103/PhysRevLett.96.140401} {\bibfield  {journal}
  {\bibinfo  {journal} {Phys. Rev. Lett.}\ }\textbf {\bibinfo {volume} {96}},\
  \bibinfo {pages} {140401} (\bibinfo {year} {2006})}\BibitemShut {NoStop}%
\bibitem [{\citenamefont {Luo}\ \emph {et~al.}(2007)\citenamefont {Luo},
  \citenamefont {Li},\ and\ \citenamefont {Bao}}]{Luo2007}%
  \BibitemOpen
  \bibfield  {author} {\bibinfo {author} {\bibfnamefont {M.}~\bibnamefont
  {Luo}}, \bibinfo {author} {\bibfnamefont {Z.}~\bibnamefont {Li}}, \ and\
  \bibinfo {author} {\bibfnamefont {C.}~\bibnamefont {Bao}},\ }\href {\doibase
  10.1103/PhysRevA.75.043609} {\bibfield  {journal} {\bibinfo  {journal} {Phys.
  Rev. A}\ }\textbf {\bibinfo {volume} {75}},\ \bibinfo {pages} {043609}
  (\bibinfo {year} {2007})}\BibitemShut {NoStop}%
\bibitem [{\citenamefont {Xu}\ \emph {et~al.}(2009)\citenamefont {Xu},
  \citenamefont {Zhang},\ and\ \citenamefont {You}}]{Xu2009}%
  \BibitemOpen
  \bibfield  {author} {\bibinfo {author} {\bibfnamefont {Z.~F.}\ \bibnamefont
  {Xu}}, \bibinfo {author} {\bibfnamefont {Y.}~\bibnamefont {Zhang}}, \ and\
  \bibinfo {author} {\bibfnamefont {L.}~\bibnamefont {You}},\ }\href {\doibase
  10.1103/PhysRevA.79.023613} {\bibfield  {journal} {\bibinfo  {journal} {Phys.
  Rev. A}\ }\textbf {\bibinfo {volume} {79}},\ \bibinfo {pages} {023613}
  (\bibinfo {year} {2009})}\BibitemShut {NoStop}%
\bibitem [{\citenamefont {Xu}\ \emph {et~al.}(2010{\natexlab{a}})\citenamefont
  {Xu}, \citenamefont {Zhang}, \citenamefont {Zhang},\ and\ \citenamefont
  {You}}]{Xu2010}%
  \BibitemOpen
  \bibfield  {author} {\bibinfo {author} {\bibfnamefont {Z.~F.}\ \bibnamefont
  {Xu}}, \bibinfo {author} {\bibfnamefont {J.}~\bibnamefont {Zhang}}, \bibinfo
  {author} {\bibfnamefont {Y.}~\bibnamefont {Zhang}}, \ and\ \bibinfo {author}
  {\bibfnamefont {L.}~\bibnamefont {You}},\ }\href {\doibase
  10.1103/PhysRevA.81.033603} {\bibfield  {journal} {\bibinfo  {journal} {Phys.
  Rev. A}\ }\textbf {\bibinfo {volume} {81}},\ \bibinfo {pages} {033603}
  (\bibinfo {year} {2010}{\natexlab{a}})}\BibitemShut {NoStop}%
\bibitem [{\citenamefont {Shi}(2010)}]{Shi2010}%
  \BibitemOpen
  \bibfield  {author} {\bibinfo {author} {\bibfnamefont {Y.}~\bibnamefont
  {Shi}},\ }\href {\doibase 10.1103/PhysRevA.82.023603} {\bibfield  {journal}
  {\bibinfo  {journal} {Phys. Rev. A}\ }\textbf {\bibinfo {volume} {82}},\
  \bibinfo {pages} {023603} (\bibinfo {year} {2010})}\BibitemShut {NoStop}%
\bibitem [{\citenamefont {Zhang}\ \emph {et~al.}(2010)\citenamefont {Zhang},
  \citenamefont {Xu}, \citenamefont {You},\ and\ \citenamefont
  {Zhang}}]{Zhang2010}%
  \BibitemOpen
  \bibfield  {author} {\bibinfo {author} {\bibfnamefont {J.}~\bibnamefont
  {Zhang}}, \bibinfo {author} {\bibfnamefont {Z.~F.}\ \bibnamefont {Xu}},
  \bibinfo {author} {\bibfnamefont {L.}~\bibnamefont {You}}, \ and\ \bibinfo
  {author} {\bibfnamefont {Y.}~\bibnamefont {Zhang}},\ }\href {\doibase
  10.1103/PhysRevA.82.013625} {\bibfield  {journal} {\bibinfo  {journal} {Phys.
  Rev. A}\ }\textbf {\bibinfo {volume} {82}},\ \bibinfo {pages} {013625}
  (\bibinfo {year} {2010})}\BibitemShut {NoStop}%
\bibitem [{\citenamefont {Xu}\ \emph {et~al.}(2010{\natexlab{b}})\citenamefont
  {Xu}, \citenamefont {Mei}, \citenamefont {L\"u},\ and\ \citenamefont
  {You}}]{Xu2010b}%
  \BibitemOpen
  \bibfield  {author} {\bibinfo {author} {\bibfnamefont {Z.~F.}\ \bibnamefont
  {Xu}}, \bibinfo {author} {\bibfnamefont {J.~W.}\ \bibnamefont {Mei}},
  \bibinfo {author} {\bibfnamefont {R.}~\bibnamefont {L\"u}}, \ and\ \bibinfo
  {author} {\bibfnamefont {L.}~\bibnamefont {You}},\ }\href {\doibase
  10.1103/PhysRevA.82.053626} {\bibfield  {journal} {\bibinfo  {journal} {Phys.
  Rev. A}\ }\textbf {\bibinfo {volume} {82}},\ \bibinfo {pages} {053626}
  (\bibinfo {year} {2010}{\natexlab{b}})}\BibitemShut {NoStop}%
\bibitem [{\citenamefont {Xu}\ \emph {et~al.}(2011)\citenamefont {Xu},
  \citenamefont {L\"u},\ and\ \citenamefont {You}}]{Xu2011}%
  \BibitemOpen
  \bibfield  {author} {\bibinfo {author} {\bibfnamefont {Z.~F.}\ \bibnamefont
  {Xu}}, \bibinfo {author} {\bibfnamefont {R.}~\bibnamefont {L\"u}}, \ and\
  \bibinfo {author} {\bibfnamefont {L.}~\bibnamefont {You}},\ }\href {\doibase
  10.1103/PhysRevA.84.063634} {\bibfield  {journal} {\bibinfo  {journal} {Phys.
  Rev. A}\ }\textbf {\bibinfo {volume} {84}},\ \bibinfo {pages} {063634}
  (\bibinfo {year} {2011})}\BibitemShut {NoStop}%
\bibitem [{\citenamefont {Shi}\ and\ \citenamefont {Ge}(2011)}]{Shi2011}%
  \BibitemOpen
  \bibfield  {author} {\bibinfo {author} {\bibfnamefont {Y.}~\bibnamefont
  {Shi}}\ and\ \bibinfo {author} {\bibfnamefont {L.}~\bibnamefont {Ge}},\
  }\href {\doibase 10.1103/PhysRevA.83.013616} {\bibfield  {journal} {\bibinfo
  {journal} {Phys. Rev. A}\ }\textbf {\bibinfo {volume} {83}},\ \bibinfo
  {pages} {013616} (\bibinfo {year} {2011})}\BibitemShut {NoStop}%
\bibitem [{\citenamefont {Xu}\ \emph {et~al.}(2012)\citenamefont {Xu},
  \citenamefont {Wang},\ and\ \citenamefont {You}}]{Xu2012}%
  \BibitemOpen
  \bibfield  {author} {\bibinfo {author} {\bibfnamefont {Z.~F.}\ \bibnamefont
  {Xu}}, \bibinfo {author} {\bibfnamefont {D.~J.}\ \bibnamefont {Wang}}, \ and\
  \bibinfo {author} {\bibfnamefont {L.}~\bibnamefont {You}},\ }\href {\doibase
  10.1103/PhysRevA.86.013632} {\bibfield  {journal} {\bibinfo  {journal} {Phys.
  Rev. A}\ }\textbf {\bibinfo {volume} {86}},\ \bibinfo {pages} {013632}
  (\bibinfo {year} {2012})}\BibitemShut {NoStop}%
\bibitem [{\citenamefont {Li}\ \emph {et~al.}(2015)\citenamefont {Li},
  \citenamefont {Zhu}, \citenamefont {He}, \citenamefont {Wang}, \citenamefont
  {Guo}, \citenamefont {Xu}, \citenamefont {Zhang},\ and\ \citenamefont
  {Wang}}]{Li2015}%
  \BibitemOpen
  \bibfield  {author} {\bibinfo {author} {\bibfnamefont {X.}~\bibnamefont
  {Li}}, \bibinfo {author} {\bibfnamefont {B.}~\bibnamefont {Zhu}}, \bibinfo
  {author} {\bibfnamefont {X.}~\bibnamefont {He}}, \bibinfo {author}
  {\bibfnamefont {F.}~\bibnamefont {Wang}}, \bibinfo {author} {\bibfnamefont
  {M.}~\bibnamefont {Guo}}, \bibinfo {author} {\bibfnamefont {Z.-F.}\
  \bibnamefont {Xu}}, \bibinfo {author} {\bibfnamefont {S.}~\bibnamefont
  {Zhang}}, \ and\ \bibinfo {author} {\bibfnamefont {D.}~\bibnamefont {Wang}},\
  }\href {\doibase 10.1103/PhysRevLett.114.255301} {\bibfield  {journal}
  {\bibinfo  {journal} {Phys. Rev. Lett.}\ }\textbf {\bibinfo {volume} {114}},\
  \bibinfo {pages} {255301} (\bibinfo {year} {2015})}\BibitemShut {NoStop}%
\bibitem [{\citenamefont {Trautmann}(2016)}]{ArnoTrautmann2016}%
  \BibitemOpen
  \bibfield  {author} {\bibinfo {author} {\bibfnamefont {A.}~\bibnamefont
  {Trautmann}},\ }\href {\doibase 10.11588/heidok.00021712} {Ph.D. thesis},\
  \bibinfo  {school} {Spin Dynamics and Feshbach Resonances in Ultracold
  Sodium-Lithium Mixtures, Combined Faculties for the Natural Sciences and for
  Mathematics of the Ruperto-Carola University of Heidelberg, Germany}
  (\bibinfo {year} {2016})\BibitemShut {NoStop}%
\end{thebibliography}%

\end{document}